\documentclass[aps,final,reprint,twocolumn,floatfix,notitlepage]{revtex4-1}
\usepackage{graphicx,amssymb,soul,color,amsmath,bm}
\usepackage{epstopdf,hyperref}
\usepackage{braket,dsfont}
\usepackage[mathcal]{euscript}

\hypersetup{colorlinks=true,citecolor=blue,linkcolor=magenta}

\sethlcolor{yellow}
\allowdisplaybreaks

\newcommand\vsigma[0]{\ensuremath{{\vec{\sigma}}}}
\newcommand\valpha[0]{\ensuremath{{\vec{\alpha}}}}

\begin{document}
\title{A machine learning approach to dynamical properties of quantum many-body systems}
\author{Douglas Hendry}
\affiliation{Department of Physics, Northeastern University, Boston, Massachusetts 02115, USA}
\author{Adrian E. Feiguin}
\affiliation{Department of Physics, Northeastern University, Boston, Massachusetts 02115, USA}

\date{\today}
\begin{abstract}
Variational representations of quantum states abound and have successfully been used to guess ground-state properties of quantum many-body systems. Some are based on partial physical insight (Jastrow, Gutzwiller projected, and fractional quantum Hall states, for instance), and others operate as a black box that may contain information about the underlying structure of entanglement and correlations (tensor networks, neural networks) and offer the advantage of a large set of variational parameters that can be efficiently optimized. However, using variational approaches to study excited states and, in particular, calculating the excitation spectrum, remains a challenge. We present a variational method to calculate the dynamical properties and spectral functions of quantum many-body systems in the frequency domain, where the Green's function of the problem is encoded in the form of a restricted Boltzmann machine (RBM). We introduce a natural gradient descent approach to solve linear systems of equations and use Monte Carlo to obtain the dynamical correlation function. In addition, we propose a strategy to regularize the results that improves the accuracy dramatically. As an illustration, we study the dynamical spin structure factor of the one dimensional $J_1-J_2$ Heisenberg model. The method is general and can be extended to other variational forms.
\end{abstract}

\maketitle
\section{Introduction}

In the past couple of years, machine learning has permeated many areas of physics and found numerous applications in condensed matter physics and chemistry. These ideas acquire a very special meaning in the context of the quantum many-body problem where one deals with datasets that are exponentially large. Sophisticated techniques have been developed to tackle this difficult challenge, such as compressing the data by using information theory and machine learning tools \cite{Freericks2014} very similar in spirit to algorithms to compress images and videos. 
In our case, datasets are comprised of all possible electronic configurations and cannot be stored in the memory of the largest supercomputer. This is an ``extreme data science'' problem from an information processing perspective, and can be approached by means of importance sampling using stochastic methods such as Monte Carlo (MC) techniques. This process can be greatly simplified if one recognizes complex patterns in the data, which has led to a line of research now called quantum machine learning\cite{Biamonte2017} that uses machine-learning algorithms to extract insightful information about quantum systems. 

Even though novel approaches based on tensor networks \cite{Orus2014} hold promise for developing efficient and accurate algorithms to solve two-dimensional(2D) many-body problems, the density matrix renormalization group (DMRG) \cite{White1992,White1993,Schollwock2005,Schollwock2011,Feiguin2013a} method has remained as the method of choice. Although understood in the context of quantum information theory, these methods share the same underlying structure and are strongly rooted on machine learning ideas such as the low rank approximation behind principal component analysis (PCA).
However, despite the success of DMRG for one-dimensional(1D) and quasi-one-dimensional geometries, extensions to actual two-dimensional systems remain challenging and applications are constrained to long cylinders and strips. The main hurdle is the fact that the number of states required to accurately represent a quantum many-body state is determined by the behavior of the entanglement entropy, the so-called ``area law''. 

Neural networks have successfully been used as variational wave function approximators to model the ground state of many-body quantum systems. The most promising results so far were achieved with restricted Boltzmann machines (RBM) \cite{Carleo2017,Saito2017,Cai2018,Glasser2018}. 
RBMs are a type of artificial neural network which are widely used in machine learning to model the probability distribution of a given data set of binary vectors drawn from an unknown probability distribution.  The components of these vectors comprise the visible layer of the neural network.  In addition to the visible layer, one introduces a hidden layer which corresponds to the components of another set of binary vectors.  These hidden vectors are auxiliary variables that expand the space of parameters and are ultimately factored out. The probability distribution of the visible vectors is formulated by first introducing a joint probability distribution for pairs of visible and hidden vectors from an energy function and Boltzmann weighting.  Finally, the probability distribution for visible vectors is taken to be the sum of the joint probability distribution over all possible configurations of the hidden vectors:
Carleo and Troyer\cite{Carleo2017} introduced a variational wavefunction for a spin-$\frac{1}{2}$ system of $N$ sites, which is inspired by the functional form of RBM.  The visible layer corresponds to the spin configurations $\vsigma^z=(\sigma^z_1,\sigma^z_2,\cdots,\sigma^z_N)$. Then the coefficients of the wave function $|\psi\rangle = \sum_{\vsigma^z} \psi(\vsigma^z)|\sigma^z_1,\sigma^z_2,\cdots,\sigma^z_N\rangle$ are represented as:
\[
\psi(\vsigma^z,\vec{a},\vec{b},W) = \sum_{h_1,h_2,\cdots,h_M} e^{-E(\vsigma,\vec{h})},
\]
with 
-\[
-E(\vsigma,\vec{h})= {\sum_{i=1}^N a_i \sigma^z_i + \sum_{i=1}^N b_i h_i +\sum_{i=1}^N\sum_{j=1}^M W_{ij} \sigma^z_i h_j},
\]
where $h_i \in \{-1,1\}$ are hidden spin variables, and $a_i,b_i,W_{ij}$ are the weights. The terms in the exponents then correspond to the negative energies for pairs of hidden and visible vectors and determine the coefficients of the wave function. The range of values the variational wave function can take on increases as the number of the hidden spin variables $M$ increases.  The summation over hidden layer vectors can be factored out which reduce the wave functions coefficients $\psi(\vsigma^z,\vec{a},\vec{b},W) = e^{\sum_{i=1}^N a_i \sigma^z_i }\prod_{j=1}^M 2\cosh{(\theta_j)}$ where $\theta_j = b_j + \sum_{i=1}^N W_{ij}\sigma^z_i$.

We propose to generalize this approach to the calculation of excited states. A simple naive idea would be to utilize $(H-\omega)^2$ as the new Hamiltonian, where $\omega$ is the target energy. However, we will take an unconventional route that will shield more valuable information: the spectral function of the problem. 

 The knowledge of the excitation spectrum of a system allows for direct
comparison with experiments, such as photoemission, or neutron scattering,
for instance.  The numerical evaluation of dynamical correlation
functions remains a very difficult task, since most computational methods are usually
capable of calculating the ground-state and maybe some low energy excitations.
A number of techniques have been used
in the past: exact diagonalization\cite{Dagotto1994} is limited to small clusters,
quantum Monte Carlo suffers from the sign problem, and requires uncontrolled analytic continuations and the use of the max entropy approximation\cite{Schuttler1986,Sandvik1998,Silver1990,Gubernatis1991,Syljuaasen2008,Fuchs2010,Sandvik2016,Shao2017}, and dynamical DMRG\cite{Hallberg1995,Kuhner1999,Jeckelmann2002,Nocera2016} is computationally very expensive. The time-dependent density matrix renormalization group and recent variations using Chebyshev expansions have been important developments, giving access to accurate spectra for very large one-dimensional systems \cite{Daley2004,White2004a,Feiguin2005,vietri,Feiguin2013b,Holzner2011,Wolf2015,Xie2018}. Matrix product states can also be used to propose variational forms for excited states\cite{Vanderstraeten2015,Vanderstraeten2015b}. Similar ideas were explored with variational Monte Carlo, that can be easily extended to higher dimensions and are free from the sign problem\cite{Li2010,DallaPiazza2014,Ferrari2018}. 

The method we introduce is derived from the so-called dynamical DMRG\cite{Jeckelmann2002} (DDMRG) and correction vector DMRG\cite{Kuehner1999}.
We plan to extract the entire dynamics of the problem by calculating the Green's function
\[
G_{ij}(z)=\langle\psi| A^\dagger_i \frac{1}{z-\hat{H}} A_j|\psi\rangle
\]
where $A$ is some operator of interest and $z=\omega+E_0+i\eta$. We derive an optimization approach based on quantum geometry concepts that will allow us to solve a large system of equations stochastically with RBMs. The method is described in great detail in sec. \ref{section:method} and we present results for the frustrated Heisenberg chain in sec. \ref{section:results}. We finally close with a discussion.  

\section{Method}\label{section:method}

\subsection{Variational solution}\label{section:VMC}

The variational wave-function $|\psi(\alpha_1,\alpha_2,\cdots)\rangle$ is parametrized by a number $N_\alpha$ of coefficients $\valpha$. In the case of an RBM, $\valpha$ represents set entire set of parameters ${\vec{a},\vec{b},W}$. 
The variational calculation of the ground state is carried out by minimizing the energy functional:
\begin{equation}
E_{var}(\valpha^*,\valpha)=\frac{\langle \psi_{\valpha^*} |\hat{H}|\psi_\valpha\rangle}{\langle \psi_{\valpha^*}|\psi_\valpha \rangle}
\label{Evar}
\end{equation}
with respect to the variational parameters $\valpha$.
A serious difficulty that plagues these calculations is the optimization procedure, due to the fact that: (i) the space of configurations grows exponentially with the number of spins; (ii) the number of model parameters to be optimized increases quadratically with the system size, making calculations prone to be trapped in local metastable solutions.
Since the number of configurations is exponentially large, the estimators are carried out by means of variational Monte Carlo. For this purpose, the variational energy is recast as:
\[
E_{var}(\valpha^*,\valpha)=\sum_{\vsigma}P_\valpha(\vsigma) E_{loc}(\vsigma),
\]
where the sum runs over all possible spin configurations $\vsigma$ and
\[
P_\valpha(\vsigma) = \frac{|\psi_\valpha(\vsigma)|^2} {\sum_{\sigma'}|\psi_\valpha(\vsigma')|^2}; \,\,\, E_{loc}(\vsigma) = \frac{\langle \vsigma|H|\psi \rangle}{\psi_\valpha(\vsigma)}
\]
with $\psi_\valpha(\vsigma)=\langle \vsigma|\psi(\valpha) \rangle$ (we omit for now the $z$ superscript, since these considerations are generic and the variables $\vsigma$ may represent arbitrary degrees of freedom).
The quantity $P(\vsigma)$ has the properties of a probability distribution, {\it{i.e}}, it is positive and normalized. This enables us to carry out a stochastic sampling of spin configurations according to $P$. In practice, since $S^z_{Tot}$ is conserved, we generate new states by randomly picking a pair of anti-parallel spins, and accepting or rejecting the new configuration with a transition probability $w=\min{(1,P_{new}/P_{old})}$. The expectation value of an observable such as the energy is then obtained by averaging over all the sampled configurations $\langle \hat{O} \rangle = \frac{1}{N_c} \sum_n^{N_c} \langle \vsigma_n|\hat{O}|\vsigma_n\rangle$. This process can be efficiently parallelized, with many Markov chains running simultaneously on different threads.

\subsection{Wave function optimization}\label{section:SR}

The number of variational parameters typically grows extensively with system size as $L$, or as $L^2$, translating into a very complex energy landscape $E_{var}(\valpha)$ with many local maxima/minima, and one global minimum that we seek. Many minimization/optimization methods can be found in the literature\cite{Harju1997,Sorella2000,Umrigar2005,Sorella2005} and here we settle for the so-called Stochastic Reconfiguration (SR) \cite{Sorella1998, Sorella2000,Sorella2005} with the optimizations proposed in Ref.\onlinecite{Neuscamman2012}. We refer the reader to a pedagogical description in Ref.\cite{Glasser2018}, that we summarize and extend here for completeness and future reference using the concept of ``natural gradient descent'' \cite{Amari1998} (NGD) (both concepts, SR and NGD, are equivalent). 

Solving for the variational parameters using Euclidean gradient descent results in each $\alpha_i$ being updated iteratively as 
$$ \alpha_i \rightarrow \alpha_i - \tau \frac{\partial E}{\partial \alpha_i^*}, $$
where $\tau$ is a small number (the ``learning rate'').
$E_{var}$ and its derivatives are estimated by sampling over states:
\begin{equation}
f_i = \frac{\partial E_{var}}{\partial \alpha_i^*}  = \langle \hat{O}_i^\dagger \hat{H} \rangle - \langle \hat{O}_i^\dagger \rangle\langle \hat{H} \rangle.
\label{forces}
\end{equation}
where the operators $\hat{O}_i$ are formally defined as the log derivatives:
\[
\hat{O}_i=\frac{1}{\psi_\valpha}\frac{\partial \psi_\valpha}{\partial \alpha_i}.
\]

By approximating the ground state as a variational wave function, we are restricting our possible wave functions to a sub-manifold of the overall Hilbert space.  This sub-manifold will be in general highly non-linear and thus have varying curvature in different directions of $\valpha$.  This can cause Euclidean gradient descent to have poor convergence.  

In order to account for this particular geometry, we utilize natural gradient descent. The gradient of a function is dependent on the metric of its domain. The vector of partial derivatives is only the gradient for the Euclidean metric (metric tensor equal to the identity).  For a non-Euclidean metric, the gradient is obtained by multiplying the inverse of the metric tensor to the vector of partial derivatives ({\it i.e.}, the Euclidean gradient). The basic idea behind NGD is to carry out gradient descent with the metric corrected gradient\cite{Sorella2010}.  

 Since the variational parameters $\alpha$ map to points on a sub-manifold of the Hilbert space, we use the metric imposed by this Hilbert space, which is the Fubini-Study metric \cite{Provost1980,Brody2001}, with distance between wave functions $|\psi\rangle$ and $|\phi\rangle$  given by 
$$\gamma(\psi,\phi)=\arccos{ \sqrt{ \frac{\langle\psi|\phi\rangle\langle\phi|\psi\rangle}{ \langle\psi|\psi\rangle\langle\phi|\phi\rangle }} }$$.
 This distance accounts for the fact that the Hilbert space is a projective space: wave functions that differ only by magnitude or an overall phase are equivalent.  
Solving for a distance would be an unnecessary constraint on the problem which is already constrained by the variational representation of the wave function.
In differentiable form, the Fubini-Study metric is given by.
        $$ds^2=\gamma(\psi,\psi+\delta\psi)^2=\frac{\langle\delta\psi|\delta\psi\rangle}{\langle\psi|\psi\rangle}-\frac{\langle\delta\psi|\psi\rangle}{\langle\psi|\psi\rangle}\frac{\langle\psi|\delta\psi\rangle}{\langle\psi|\psi\rangle}$$
        Using this we can calculate the induced metric tensor on our variational parameters by equating the differentiable distances given by the metric tensor on $\valpha$ and the differentiable distances in the Hilbert space of the wave functions they map to:
        $$ds^2=\sum_{ij}{\delta\alpha_i^* g_{ij} \delta\alpha_{j}}=\gamma(\psi(\alpha),\psi(\alpha+\delta\alpha))^2 $$.
        The solution is given by:
        $$g_{ij}=\frac{\langle\partial_i\psi|\partial_{j}\psi\rangle}{\langle\psi|\psi\rangle}-\frac{\langle\partial_i\psi|\psi\rangle}{\langle\psi|\psi\rangle}\frac{\langle\psi|\partial_{j}\psi\rangle}{\langle\psi|\psi\rangle}$$,
 where $|\partial_{i}\psi\rangle=\frac{\partial}{\partial\alpha_i}|\psi\rangle$
        Reformulating this matrix in terms of sampling over states $\vsigma$ results in the covariance matrix of the log deratives $O_i$
\begin{equation}
g_{ij} = \langle \hat{O}_i^\dagger \hat{O}_j \rangle - \langle \hat{O}_i^\dagger \rangle \langle \hat{O}_j \rangle.
\label{covariance}
\end{equation} 
        Now, at each iteration the change in variational parameters is given by solving the system of equations
\begin{equation}
\sum_{j}{g_{ij}\Delta\alpha_i}=-\tau\frac{\partial E}{\partial \alpha_i^*}.
\label{sr2}
\end{equation}

The optimization procedure consists of calculating the forces $f_i$ in (\ref{forces}) and the covariance matrix $g_{ij}$ in (\ref{covariance}) and solving (\ref{sr2}). This is carried out iteratively until converged.  
In practice, we follow the accelerated SR method proposed in Ref.\cite{Neuscamman2012}, where it is shown that the construction and storage of the matrix $g$ can be bypassed, translating into a remarkable speedup.

\subsection{Correction vector}\label{section:CV}

In order to calculate the Green's function:
\[
G_{ij}(z)=\langle\psi| A^\dagger_i \frac{1}{z-\hat{H}} A_j|\psi\rangle
\]
where $A$ is some operator of interest and $z=E_0+\omega+i\eta$ we follow a procedure pioneered in the context of matrix product states, known as dynamical DMRG\cite{Kuehner1999, Jeckelmann2002}. It requires the calculation of the following auxiliary states:
\begin{eqnarray}
|A_i\rangle &=& \hat{A}_i |\psi\rangle \nonumber \\
|\chi_j(z)\rangle &=& \frac{1}{z-\hat{H}} |A_j\rangle, 
\label{correction}
\end{eqnarray}
where $|\chi_j(z)\rangle$ is called the ``correction vector''. 
Explicitly, $|\chi_j(z)\rangle$ can be  obtained by solving the equation:
\begin{eqnarray}
(z-\hat{H})|\chi_j(z)\rangle = \hat{A}_j|\psi\rangle = |A_j\rangle.
\label{G}
\end{eqnarray}
The spectral function is defined as the imaginary part of the Green's function, $A_{ij}(\omega)=-\frac{1}{\pi}\mathrm{Im}{G_{ij}(z)}$, or:
\begin{equation}
A_{ij}(\omega)=-\frac{1}{\pi}\mathrm{Im}\langle A_i|\chi_j(z)\rangle.
\label{A_from_G}
\end{equation}
By Fourier transforming the spatial dependence to momentum, one obtains the entire excitation spectrum of the problem resolved in both momentum and frequency.

An alternative way to solve for $A_{ij}$ is to directly target the imaginary part:
\begin{eqnarray}
\left((\hat{H}-E_0-\omega)^2+\eta^2\right)|\tilde{\chi}_j\rangle = -\eta|A_j\rangle,
\label{A}
\end{eqnarray}
such that 
\begin{equation}
A_{ij}(\omega)=-\frac{1}{\pi}\langle A_i|\tilde{\chi}_j\rangle
\label{ImG}
\end{equation}
However, this form is typically more unstable: if $|\tilde{\chi}_j\rangle$ is close to an eigenstate of the Hamiltonian, the quantity in the square can become very small when approaching a pole in the spectrum. 
The fact that we are dealing with variational wave-functions implies that the results obtained by these two approaches will not necessarily be the same. We discuss the consequences below.

\subsection{Solving for the Green's function}

As previously discussed, 
 we need to solve the following system of equations for $|\chi_j(z)\rangle$
\begin{equation} (z-\hat{H})|\chi_j(z) \rangle=|A_j\rangle,\end{equation}
where $|A_j\rangle=\hat{A_j}|\psi\rangle$, and $|\chi_j(z)\rangle$ is parametrized by another set of variational parameters $\valpha$. The Green's function can then be obtained as $G_{ij}(z)=\langle A_i|\chi_j (z) \rangle$.
We hereby introduce a similar natural gradient descend procedure to the one outlined in section  \ref{section:SR} to solve the generic system of equations $\hat{Q}|\chi \rangle=|A\rangle$
(we later will apply this method to the particular case with $\hat{Q}=z-\hat{H}$).


We solve for $|\chi \rangle$ by first minimizing the Fubini-Study metric between $\hat{Q}|\chi \rangle$ and $|A\rangle$ :
\begin{equation} \ \gamma(\hat{Q}\chi,A)=\arccos{\sqrt{x}} \end{equation}
\begin{equation} \
x= \frac{ \langle \chi|\hat{Q}^\dagger |A\rangle \langle A|\hat{Q}|\chi \rangle   }{  \langle \chi|\hat{Q}^\dagger\hat{Q}|\chi \rangle \langle A|A \rangle}.
\label{angle}
\end{equation}
As discussed above, the NGD method will yield a state $\hat{Q}|\chi \rangle$ that is parallel (or as parallel as possible), to $|A\rangle$, but with unconstrained phase and norm. Therefore, the resulting wavefunction $|\chi\rangle$ is not quite the one we seek, but it is off by a constant $|\chi\rangle = \beta|\tilde{\chi}\rangle$, (where $|\tilde{\chi}\rangle$ is the actual solution) that can readily be obtained. 

 The derivatives of $\gamma^2$ are given by:
\begin{equation} \ 
\frac{\partial\gamma^2}{\partial\alpha_i^*}=\gamma\sqrt{\frac{x}{1-x}}\left[ \frac{\langle \partial_i \chi|\hat{Q}^\dagger\hat{Q}|\chi \rangle}{\langle \chi|\hat{Q}^\dagger\hat{Q}|\chi \rangle}  -\frac{\langle \partial_i \chi|\hat{Q}^\dagger|A \rangle}{\langle \chi|\hat{Q}^\dagger|A \rangle}    \right],
\label{gamma}
\end{equation}
with
\begin{equation} 
|\partial_i \chi \rangle=\sum_{\vsigma}{\frac{\partial \chi(\vsigma)}{\partial \alpha_i}|\vsigma \rangle }.
\end{equation}

The parameters $\valpha$ are updated at each iteration using stochastic reconfiguration / natural gradient descent which gives 
\begin{equation} \ 
\sum_{j}{g_{ij}\Delta \alpha_{j}}= - \lambda \frac{\partial\gamma^2}{\partial\alpha_i^*}
\label{sij}
\end{equation}
where $\lambda$ is the learning rate (a small number) and the metric tensor $g$ is derived from $\hat{Q}|\chi\rangle$(rather than $|\chi\rangle$) and is given by  
\begin{equation} \ 
g_{ij}= \frac{\langle \partial_i \chi|\hat{Q}^\dagger\hat{Q}|\partial_{j}\chi \rangle}{\langle \chi|\hat{Q}^\dagger\hat{Q}|\chi \rangle}-\frac{\langle \partial_i \chi|\hat{Q}^\dagger\hat{Q}|\chi \rangle}{\langle \chi|\hat{Q}^\dagger\hat{Q}|\chi \rangle}    \frac{\langle  \chi|\hat{Q}^\dagger\hat{Q}|\partial_{j}\chi \rangle}{\langle \chi|\hat{Q}^\dagger\hat{Q}|\chi \rangle}.
\end{equation}

Overlaps are estimated using Monte-Carlo sampling over probability distributions $P_0(\vsigma)=|\langle \vsigma|A\rangle|^2 / \langle A|A \rangle$ and $P_1(\vsigma)= |\langle \vsigma|Q|\chi\rangle|^2 / \langle \chi|\hat{Q}^\dagger\hat{Q}|\chi \rangle $.  The two probabilities will become equivalent as the wave-function converges. However it is important to sample over both distributions to account for states that have much greater weight in one distribution than the other. For each sampled configuration $\vsigma$ the following quantities are calculated:
\begin{equation} \ 
R(\vsigma)=\frac{\langle \vsigma|\hat{Q}|\chi\rangle}{\langle \vsigma |A\rangle}
\end{equation}
\begin{equation} \ 
\mathcal{O}_i(\sigma)=\frac{\langle \vsigma|\hat{Q}|\partial_i\chi\rangle}{\langle \vsigma|\hat{Q}|\chi\rangle}.
\end{equation}

\begin{figure}
    \centering
    \includegraphics[width=\linewidth]{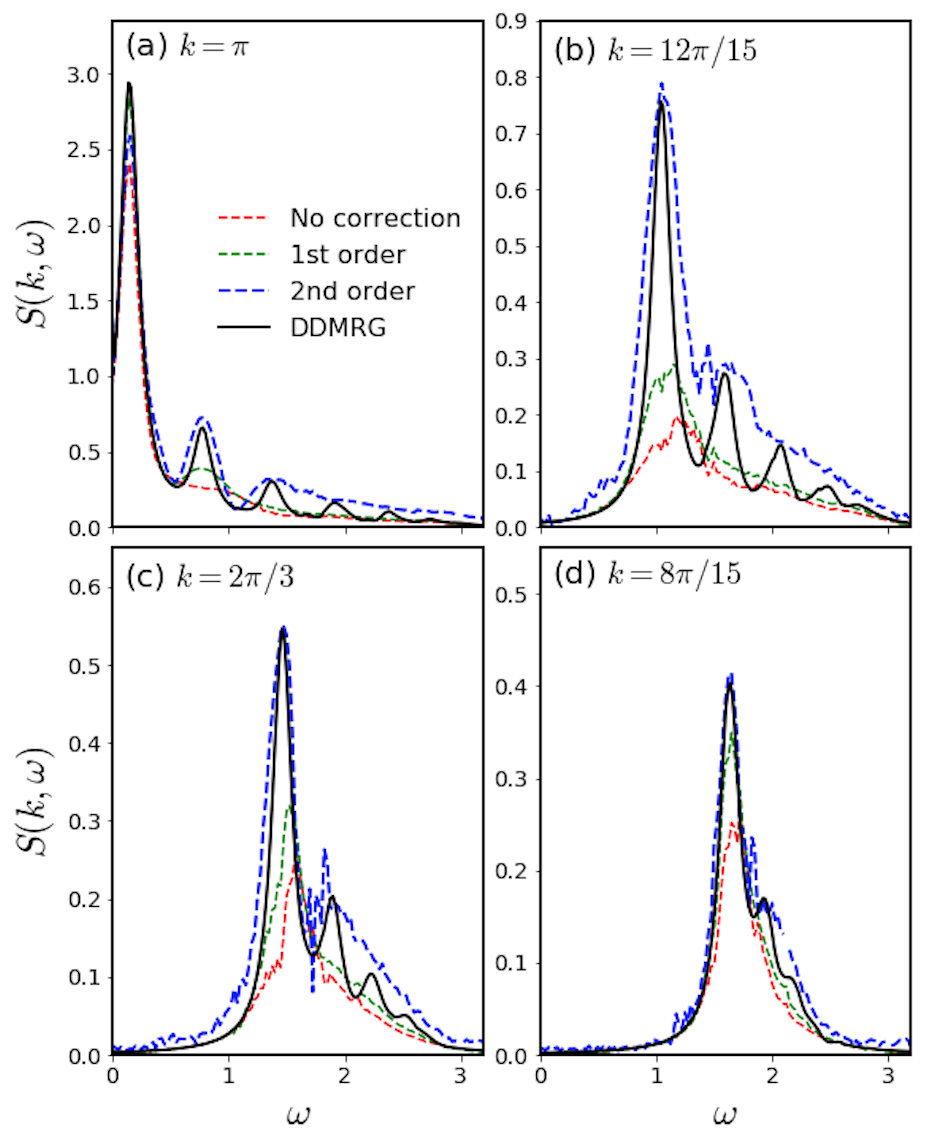}
    \caption{Spin structure factor of a Heisenberg chain of size $L=30$ with periodic boundary conditions for some representative values of momentum $k$. We show results obtained without error correction, and 1st and 2nd order regularizations. Dynamical DMRG data is also included for comparison. An artificial broadening $\eta=0.1$ was introduced in all cases. Notice that change of scale between panels in the $y$-axis. }
    \label{fig:results}
\end{figure}

Then, equations (\ref{angle}),(\ref{gamma}), and (\ref{sij}) can be expressed in terms of sampling as
\begin{eqnarray}  
x & = & \frac{|\langle R(\vsigma) \rangle_0|^2}{\langle|R(\vsigma)|^2\rangle_0} \nonumber \\
\frac{\partial\gamma^2}{\partial\alpha_i^*} &= &\gamma\sqrt{\frac{x}{1-x}}\left[ \langle\mathcal{O}_i^*(\vsigma)\rangle_1 - \frac{\langle\mathcal{O}_i^*(\vsigma)R^*(\vsigma)\rangle_0}{\langle R^*(\sigma) \rangle_0}   \right] \\
g_{ij} &=&\langle\mathcal{O}_i^*(\vsigma)\mathcal{O}_{j}(\vsigma)\rangle_1-\langle\mathcal{O}_i^*(\sigma)\rangle_1 \langle\mathcal{O}_{j}(\sigma)\rangle_1. \nonumber
\end{eqnarray}
Notice that $R$ becomes constant when the states become parallel and thus the sampling variance goes to zero, just as the local energies become constant when solving for the ground state.

\begin{figure}
    \centering
    \includegraphics[width=\linewidth]{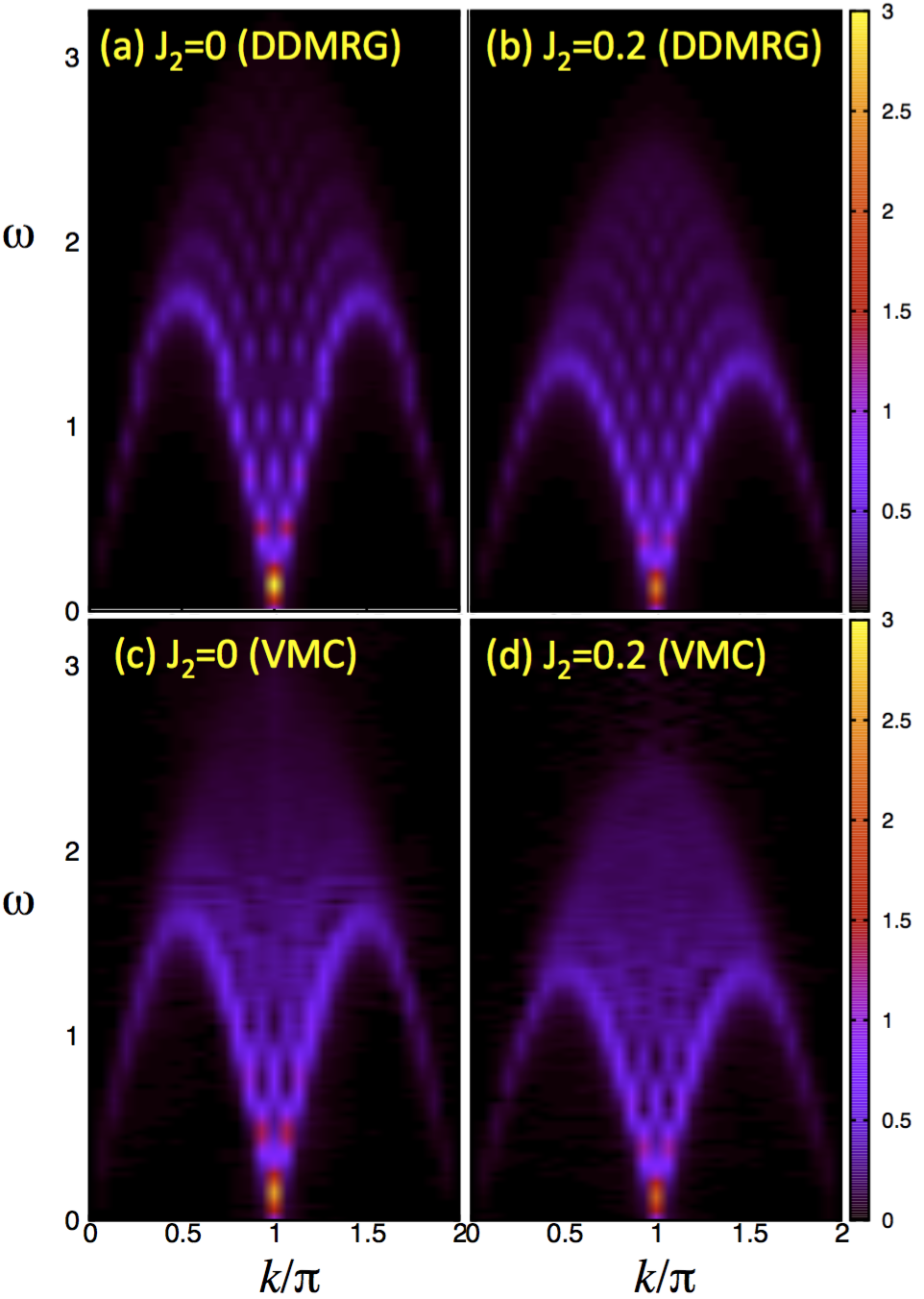}
    \caption{Momentum and frequency resolved spectra for the frustrated Heisenberg chain in the gapless regime with $J_2=0$ and $J_2=0.2$, obtained with both (a)-(b) DDMRG and (c)-(d) variational Monte Carlo using error correction.
    }
    \label{fig:results_J2}
\end{figure}

Finally, the wave function normalization (the constant $\beta$) is obtained as
$$\beta=\frac{\langle\chi|\hat{Q}^\dagger|A\rangle}{\langle\chi|\hat{Q}^\dagger\hat{Q}|\chi\rangle},$$
which in terms of sampling is given as
$$\beta=\frac{\langle R(\sigma)\rangle_0^*}{\langle|R(\sigma)|^2\rangle_0}$$

\subsection{Error correction}

The Green's function can be obtained by solving either one of the two equations:
\begin{equation} \hat{Q}|\chi^{+}\rangle =|A\rangle \end{equation}
or
\begin{equation} \hat{Q}^\dagger|\chi^{-}\rangle =|A\rangle, \end{equation}
where $|A\rangle=\hat{A}|\psi \rangle$, $\hat{Q}=E_0+\omega+i\eta-\hat{H}$, and $\hat{Q}^\dagger=E_0+\omega-i\eta-\hat{H}$.
We can then solve for $G$ in three different ways
\begin{equation} G=-\frac{1}{\pi}\langle A|\chi^{+}\rangle=-\frac{1}{\pi}\langle \chi^{-}|A\rangle=-\frac{1}{\pi}\langle \chi^{-}|\hat{Q}|\chi^{+}\rangle \end{equation}

As we discussed earlier, the NGD method will allow us to find a wave function as close as possible to the one we seek. However, it is possible that this wave-function does not accept a faithful representation in terms of the proposed variational form. As a consequence, regardless of the sampling error, there always will be an inherent error due to the limitations of the wave function representation.
Now let $|\tilde{\chi}^{+}\rangle$ and $|\tilde{\chi}^{-}\rangle$ be the variational wave function approximations with errors $\epsilon^{+}|\phi^{+}\rangle$ and $\epsilon^{-}|\phi^{-}\rangle$ respectively so that 
\begin{eqnarray} 
|\tilde{\chi}^{+}\rangle & = & \hat{Q}^{-1}|A\rangle + \epsilon^{+}|\phi^{+}\rangle, \nonumber \\
|\tilde{\chi}^{-}\rangle  & = & (\hat{Q}^\dagger)^{-1}|A\rangle + \epsilon^{-}|\phi^{-}\rangle. 
\end{eqnarray}
Then we calculate $G$ all three ways, each with different but related error terms:
\begin{eqnarray} 
-\frac{1}{\pi}\langle A|\tilde{\chi}^{+}\rangle & = & G - \frac{1}{\pi}\epsilon^{+}\langle A|\phi^{+}\rangle, \nonumber \\
 -\frac{1}{\pi}\langle \tilde{\chi}^{-}|A\rangle & = & G - \frac{1}{\pi}\epsilon^{-}\langle \phi^{-}|A\rangle, \\
-\frac{1}{\pi}\langle \tilde{\chi}^{-}|\hat{Q}|\tilde{\chi}^{+}\rangle & = & G - \frac{1}{\pi} \left[ \epsilon^{-}\langle \phi^{-}|A\rangle + \epsilon^{+}\langle A|\phi^{+}\rangle + \right. \\ \nonumber
&+& \left.\epsilon^{+}\epsilon^{-}\langle \phi^{-}|\hat{Q}|\phi^{+}\rangle \right]. \nonumber 
\label{first_order} 
\end{eqnarray} 

Combining all three estimates we can get the first order error terms to cancel:
\begin{eqnarray} -\frac{1}{\pi} \left[\langle A|\tilde{\chi}^{+}\rangle+\langle \tilde{\chi}^{-}|A\rangle-\langle \tilde{\chi}^{-}|\hat{Q}|\tilde{\chi}^{+}\rangle \right]= \\ \nonumber 
 = G - \frac{1}{\pi}\epsilon^{+}\epsilon^{-}\langle \phi^{-}|\hat{Q}|\phi^{+}\rangle.
\end{eqnarray}

However we can improve upon this by isolating another second order error term.  To do this we calculate $\langle A|A \rangle$ also in three different ways.
\begin{eqnarray} 
\langle A|\hat{Q}|\tilde{\chi}^{+}\rangle & = & \langle A|A \rangle+\epsilon^{+}\langle A|\hat{Q}|\phi^{+}\rangle, \nonumber \\
\langle \tilde{\chi}^{-}|\hat{Q}|A\rangle & = & \langle A|A \rangle+\epsilon^{-}\langle \phi^{-}|\hat{Q}|A\rangle \\
\langle \tilde{\chi}^{-}|\hat{Q}^2|\tilde{\chi}^{+}\rangle & = & \langle A|A \rangle +\epsilon^{+}\langle A|\hat{Q}|\phi^{+}\rangle \\ \nonumber 
&+&\epsilon^{-}\langle \phi^{-}|\hat{Q}|A\rangle+ \epsilon^{+}\epsilon^{-}\langle \phi^{-}|\hat{Q}^2|\phi^{+}\rangle.
\end{eqnarray}
Combining the above equations we obtain:
\begin{eqnarray} 
\langle A|A \rangle+\langle \tilde{\chi}^{-}|\hat{Q}^2|\tilde{\chi}^{+}\rangle-\langle A|\hat{Q}|\tilde{\chi}^{+}\rangle-\langle \tilde{\chi}^{-}|\hat{Q}|A\rangle= \\ \nonumber
=\epsilon^{+}\epsilon^{-}\langle \phi^{-}|\hat{Q}^2|\phi^{+}\rangle.
\label{second_order}
\end{eqnarray}
Then, in order to get our best estimate for $G$ we multiply Eq.(\ref{second_order}) by $1/(i\eta\pi)$ and add it to Eq.(\ref{first_order}) which results in:
\begin{equation} G + \frac{1}{\pi}\left[\epsilon^{+}\epsilon^{-}\frac{1}{i\eta}\langle \phi^{-}|\hat{Q}^2|\phi^{+}\rangle-\epsilon^{+}\epsilon^{-}\langle \phi^{-}|\hat{Q}|\phi^{+}\rangle \right].
\label{error_cancelation}
\end{equation}

To understand why this makes an improvement over the estimate (\ref{first_order}) we expand the wave-functions and their errors in terms of eigenstates:
\begin{eqnarray}
\hat{Q}^{-1}|A\rangle &=&  \sum_{n}{\frac{A_n}{\Delta E_n+i\eta}}|n\rangle, \nonumber \\
\epsilon^{+}|\phi^{+}\rangle &= & \sum_{n}{\left(\frac{A_n}{\Delta E_n+i\eta}\right)\epsilon^{+}_{n}}|n\rangle, \nonumber \\
(\hat{Q}^\dagger)^{-1}|A\rangle &=& \sum_{n}{\frac{A_n}{\Delta E_n-i\eta}}|n\rangle, \nonumber \\
\epsilon^{-}|\phi^{-}\rangle &=& \sum_{n}{\left(\frac{A_n}{\Delta E_n-i\eta}\right)\epsilon^{-}_{n}}|n\rangle, \nonumber
\end{eqnarray}
where $\Delta E_n = E_0+ \omega - E_n$.  Then the error term of \ref{first_order} is 
\begin{equation}  -\frac{1}{\pi}\epsilon^{+}\epsilon^{-}\langle \phi^{-}|\hat{Q}|\phi^{+}\rangle = -\frac{1}{\pi}\sum_{n}{\left(\frac{|A_n|^2}{\Delta E_n+i\eta}\right)(\epsilon^{-}_{n})^{*}\epsilon^{+}_{n}}
\end{equation}
Multiplying the isolated error (\ref{second_order}) by $1/(i\eta\pi)$ yields: 
\begin{equation}  \frac{1}{\pi}\frac{1}{i\eta}\epsilon^{+}\epsilon^{-}\langle \phi^{-}|\hat{Q}^2|\phi^{+}\rangle = \frac{1}{\pi}\sum_{n}{\left(\frac{|A_n|^2}{i\eta}\right)(\epsilon^{-}_{n})^{*}\epsilon^{+}_{n}}.
\end{equation}
Finally, the error of Eq.(\ref{error_cancelation}) is then  
\begin{eqnarray} \frac{1}{\pi}\left[\epsilon^{+}\epsilon^{-}\frac{1}{i\eta}\langle \phi^{-}|\hat{Q}^2|\phi^{+}\rangle-\epsilon^{+}\epsilon^{-}\langle \phi^{-}|\hat{Q}|\phi^{+}\rangle \right] &=& \\ \nonumber
 = \frac{1}{\pi}\sum_{n}{\left(\frac{|A_n|^2}{i\eta}\right)\left(\frac{\Delta E_n}{\Delta E_n+i\eta}\right)(\epsilon^{-}_{n})^{*}\epsilon^{+}_{n}}.
\end{eqnarray}
The dominant eigenstate $|n\rangle$ in these expressions is the one such that $\Delta E_n\approx0$ and thus the dominant error terms should be $\epsilon^{-}_{n}$ and $\epsilon^{+}_{n}$.  But those terms are multiplied by a $\Delta E_n$ in the numerator and, as a result, we not only eliminate the first order errors terms, but also the part of the second order error from the most dominating contribution.  

\section{Results}\label{section:results}

For illustration purposes we will focus on the one-dimensional spin-$\frac{1}{2}$ Heisenberg model with nearest and next nearest neighbor interactions, the so-called $J_1-J_2$ model:
\begin{equation}
\hat{H}=\sum_{i=0}^{L-1} {\left(J_1 \vec{S}_{i}\cdot\vec{S}_{i+1} + J_2 \vec{S}_{i}\cdot\vec{S}_{i+2}\right)},
\end{equation}
where $\vec{S}=(\hat{S}^x,\hat{S}^y,\hat{S}^z)$ are spin operators. We consider periodic boundary conditions and chose $J_1$ as our unit of energy.
We calculate the spin structure factor, defined as:
\[
S^z(k,\omega) = -\frac{1}{L\pi} \mathrm{Im} \sum_n e^{ikn}{\langle \psi| \hat{S}^z_0 \frac{1}{z-\hat{H}} \hat{S}^z_n |\psi \rangle}.
\]
where we have used translational invariance, since the system has periodic boundary conditions.
In order to calculate the correlation function we solve the following system of equations for $|\psi(z,j)\rangle$ using the prescription described in the previous section:
\begin{equation} \label{eq:someequation}(z-\hat{H})|\psi (z,j) \rangle=|A_j\rangle,
\end{equation}
where $|A_j\rangle=\hat{S}^z_j|\psi_0\rangle$. Finally, $G_{ij}(z)=\langle A_i|\psi (z,j) \rangle$.

We typically carry out computations taking 20,000 measurements for each optimization step, leaving 100 iterations in between measurements to make sure they are independent and uncorrelated. We took 10$^6$ samples for the overlaps. We then solve for the wave-functions $|\chi^\pm\rangle$ and calculate the Green's functions using error correction as described in the previous section. 

We studied chains of length $L=30$, larger than the largest system achievable using exact diagonalization, but still smaller than what DMRG can solve. We used 120 hidden variables in the hidden layer that translates into $\sim 3000$ variational parameters. As a benchmark, we compare our results to dynamical DMRG calculations with $m=600$ DMRG states using a broadening $\eta=0.1$.

We first show results obtained with first order and second order error correction in Fig.\ref{fig:results} for several representative values of momentum $k$ and $J_2=0$. While the range and position of low energy poles agrees quite well, we observe a remarkable improvement upon introducing the second order correction that is particularly marked around the cusp of the peaks/poles. The range in frequency in between poles is not so accurately matched. While we observe some oscillations that we attribute to numerical errors, the main source of discrepancy s likely due to the limitations of the variational wave-function utilized. This is understood using the arguments discussed in the previous section: the second order error gets practically suppressed when the frequency corresponds to an eigenstate $\omega \sim E_n-E_0$. It is expected that as the system size $L$ increases and the spectrum becomes continuous, the errors will be practically cancelled and the accuracy will improve over the entire range of frequencies. We next show the spectrum for $J_2=0$ and $J_2=0.2J_1$ in Figs.\ref{fig:results} and \ref{fig:results_J2}, both in a color scale and frequency cuts for a couple of momenta. The width and the edge of the spinon continuum are very well described, as well as the magnitude of the excitation peaks. 

\section{Conclusions}
We have presented a variational approach to calculate Green's functions and dynamical structure factors of many-body quantum systems directly in the frequency domain using restricted Boltzmann machines. The method, inspired in dynamical DMRG and machine learning concepts, allows one to obtain the entire spectrum of excitations, which in the Heisenberg model consists of deconfined domain walls (spinons).  In order to solve for the Green's functions, we introduce a natural gradient descent method to solve complex systems of equations where the solution is encoded in RBM form. The problem is solved stochastically and can be parallelized to run different frequencies on different computing threads or nodes. Unlike the VMC method of Ferrari {\it et al.} \cite{Ferrari2018} which can provide a few hundred discrete poles, our method yields the entire spectrum with full frequency resolution. 
These ideas are not limited to a particular form of variational wave function and is completely general (DDMRG does it with matrix product states). In particular, we show that RBMs are not able to faithfully represent excited states but, nonetheless, we are able to reconstruct the spectral functions very accurately by introducing a regularization scheme that eliminates first and second order errors. We demonstrate the application of the technique to the frustrated case away from integrability, where our results accurately describe the position of the poles (especially low frequency ones) and the continuum. The approach can be naturally extended to higher dimensions, where both quantum Monte Carlo and the DMRG have shortcomings.

\acknowledgments
The authors acknowledge the National Science Foundation for support under grant No. DMR-180781.

%

\end{document}